\documentclass[letterpaper]{IEEEtran}
\usepackage[dvipsnames]{xcolor}
\usepackage{enumerate}
\usepackage{amssymb}
\usepackage{amsmath}
\usepackage{soul}
\usepackage{epsfig}
\usepackage{multirow}
\usepackage{verbatim}
\usepackage{graphicx}
\usepackage{epsfig}
\usepackage{amsmath, amssymb}
\usepackage{subfigure}
\usepackage{times}
\usepackage[ruled,linesnumbered]{algorithm2e}
\SetAlFnt{\small}
\DontPrintSemicolon
\usepackage{xcolor}
\usepackage{cite}
\usepackage{pgfplots}
\usepackage{multicol}
\usepackage{mathrsfs}
\usepackage{color}
\usepackage{tikz}
\usetikzlibrary{plotmarks,shapes,snakes}
\usepackage{pgfplots}
\usepackage{listings}
\usepackage{xspace}
\usepackage{newtxmath}
\usepackage{pdfpages}

\newcommand{\eg}{e.g.,\xspace}
\newcommand{\ie}{{\em i.e.,}\xspace}

\newcommand{\BfPara}[1]{{\noindent\bf#1.}\xspace}

\newcommand{\ours}{FlowRegulator}
\newcommand{\ourFR}{FlowRegulator}

\begin{document}
	
	\title{Scaling Up Anomaly Detection Using In-DRAM Working Set of Active Flows Table}
	\author{\IEEEauthorblockN{Rhongho Jang$^{\dagger\ddagger}$, Seongkwang Moon$^\dagger$, Youngtae Noh$^{\dagger}$, Aziz Mohaisen$^\ddagger$, and DaeHun Nyang$^\dagger$}
	
	$^\dagger$INHA University \hspace{10mm} $^\ddagger$ University of Central Florida
	}
	
	\maketitle

	\begin{abstract}
	In the zettabyte era, per-flow measurement becomes more challenging owing to the growth of both traffic volumes and the number of flows.
	Also, swiftness of detection of anomalies (\eg DDoS attack, congestion, link failure, and so on) becomes paramount.
	For fast and accurate anomaly detection, managing an accurate working set of active flows (WSAF) from massive volumes of packet influxes at line rates is a key challenge. WSAF is usually located in a very fast but expensive memory, such as TCAM or SRAM, and thus the number of entries to be stored is quite limited.
	To cope with the scalability issue of WSAF, we propose to use In-DRAM WSAF with scales, 
	and put a compact data structure called \ourFR{} in front of WSAF to compensate for DRAM's slow access time by substantially reducing massive influxes to WSAF without compromising measurement accuracy. We prototype and evaluated our system in a large scale real-world experiment (connected to monitoring port of our campus main gateway router for 113 hours, and capturing 122.3 million flows). As  one  key  application, \ourFR{} detected heavy hitters with 99.8\% accuracy.
	\end{abstract}

	\IEEEpeerreviewmaketitle	
	\def\dsp{\def\baselinestretch{1.00}\large\normalsize}
	\dsp
	\def\ffsp{\def\baselinestretch{1.000}\large\normalsize}
	\ffsp
	\interfootnotelinepenalty=10000

\vspace{-5mm}
\section{Introduction}
\label{sec:intro}
Due to the high speed network, traffic measurement now have to cope with 
enormous incoming data rates (\ie larger number of flows) with tight deadlines (\ie real-time). 
We stress that large-scale instant measurement is highly necessary for network anomaly detection. For example, if denial of service (DoS) attacks cause an influx of packets at 100 Gbps, detection delay of 100 ms will cause 1.2GB data to hit a server or a network. Therefore, to avoid large bandwidth penalties, instant anomaly detection is essential.

A working set of active flows (WSAF) is a type of cache of a full flow table, which can be found usually in TCAM (Ternary Content Addressable Memory), CAM, or sometimes SRAM for traffic monitoring.
For instanse, NetFlow uses TCAM for storing WSAF in which an entry consists of a flow ID and the counting value. However, the number of entries of the table cannot be large because those types of memories are quite expensive.
For scalability, we can put WSAF in DRAM instead of the expensive memory (\ie incentive to cost-effectiveness). 
However, there is a speed issue for In-DRAM WSAF: a packet arrival rate is too fast to handle by In-DRAM WSAF, owing to the DRAM's speed and WSAF table's hash collision.

To cope with these issues, sketch-based techniques have been greatly enhanced over several decades because sketch-based counting algorithms only require a small amount of memory to {\em encode} a large volume of traffic in real-time. 
Due to their design, however, most of {\em decode} algorithms involve hundreds of hash calculations and memory accesses from statistically mixed random blocks to obtain meaningful statistics~\cite{Huang17}.
For this reason, offline decoding in a high-performance server is commonly accepted in practice but inherently incurs huge network delay. Thus, online decoding is highly necessary for instant measurement and further timely detection.

Unfortunately, most sketch-based algorithms lack scalability and online decoding capabilities.
Our approach to solve these two problems is 1) to use a counting algorithm that can perform
online decoding and 2) to put a flow regulator before WSAF to slow down the incoming packet rate to WSAF.
To realize both ideas, we designed a highly scalable counting and flow regulating algorithm called 
\ourFR{}.
By design, instead  of  directly  inserting or updating  every  packet  of  a  flow  into  WSAF table, 
\ourFR{} (\ie a small cache buffer) retains a fraction of flow counts. 
By doing so, we can suppress frequent WSAF updates in DRAM; thereby \ourFR{} can support 
large-scale influx of flows with the use of cost-effective large DRAM.
Consequently, \ourFR{} relaxes the necessity of expensive memories (TCAM or SRAM) for maintaining large WSAF, and further enables us to build a highly scalable and fast measurement system. We conducted a real-world campus network experiment for 113 hours by connecting our prototype of \ourFR{} to a mirroring port of a main gateway router, capturing 9.11 billion packets, 122.3 million flows, and 8.5TB bytes. \ours{} successfully measured the whole L4 flows with a standard error (0.65\%) and  detected  heavy  hitters with 99.8\%  accuracy.

\section{FlowRegulator Design}\label{sec:motivation}
Our large WSAF in DRAM is in contrast to the small WSAF in TCAM (\ie industry practice). In DRAM, we can store much more flows, thereby, we do not need a remote collector for decoding.
However, the downside is that we cannot evade the ``sluggishness'' of DRAM.

{\noindent\bf \ourFR{} to relax the \{ips = pps\} constraint:} 
Instead of directly inserting or updating every flow packet into the table, we put a small buffer called \ourFR{} to retain a fraction of flow counts before WSAF.
\ourFR{} has a memory block (or a virtual vector initialized to all 0's) for every single flow, and whenever a packet comes in, the corresponding block is updated by setting a random bit of the block. 
When the block saturates (or a portion of block has set to 1's), the resulting counting fraction (we note that this is not the total size of a flow) is added up to the WSAF (\ie a hash table in DRAM). 
Because \ourFR{} retains mice flows whose sizes are lower than the saturation condition,
not all the packets are fed into WSAF, but only the packets that trigger the saturation 
condition are given to WSAF. This design will greatly reduce insert per second (ips) even under a high packet per second (pps) condition.

To develop \ourFR{}, we utilize the recyclable counter with confinement (RCC)~\cite{nyang16} that already has online decoding capability, and proven to be useful for measurement in the wireless SDN environment~\cite{Jang:2017,JangCMNN17}.
To investigate its feasibility, we have tested RCC for its rate regulation (defined as Output ips/Input pps). 
Given that access time of SRAM  is 10-20 times faster than DRAM's (and even faster with TCAM), 
RCC's rate regulation should be less than 5\%.
However, its regulation and retention capacity (the maximum number of packets in a virtual vector) are not operationally sufficient. 
Thus, it is impossible to work with RCC for building \ourFR{}.
One way to increase the rate regulation is to give RCC a larger virtual vector, but that does not expand the retention capacity. 

{\noindent\bf Two-layer design for higher rate regulation:} 
Here, our observation is that enlarging the virtual vector size increases the retention capacity just in an addictive manner, 
and thus, this is not a viable (\ie scalable) option. Instead, we designed a new counting algorithm for \ourFR{}, 
which has two layers of probabilistic counters to achieve the higher rate regulation.
Our \ourFR{} plays a key role of retaining flows (from feeding into WSAF) for a while as well as counting flows.
In the two-layer design, the second (higher) layer's one bit encodes multiple packets of a flow
from a saturated sketch of the first (lower) layer. This design has substantially improved the rate regulation
in a multiplicative manner. It enables higher rate regulation while not being detrimental to the accuracy and speed, while being scalable.

\begin{figure}[t]
	\centering
	\includegraphics[width=0.49\textwidth]{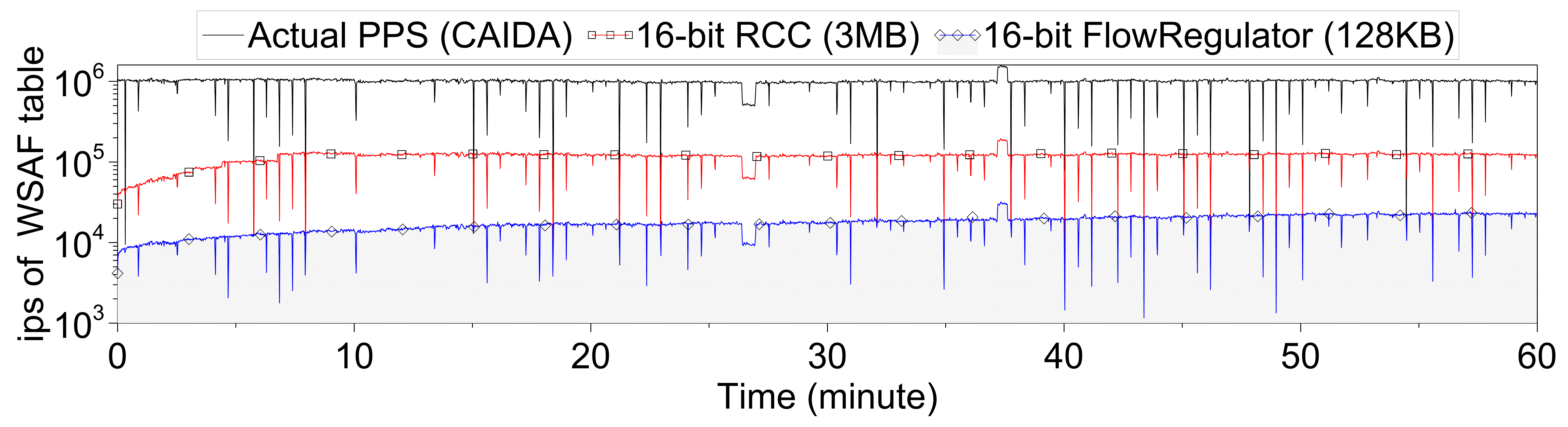}\vspace{-3mm}
	\caption{WASF relaxation: \ourFR{} (FR) and RCC ips of CAIDA dataset} \vspace{-4mm}
	\label{fig-relax}
\end{figure}
\vspace{-3mm}
\section{Evaluation} \label{sec:impl}

\BfPara{WASF ips relaxation} 
In Fig.~\ref{fig-relax}, the x-axis represents the time line of our CAIDA dataset, 
and the black solid line on the top represents the actual pps of the trace. 
Below the pps line, RCC's and \ourFR{}'s regulation rates are shown in red squares and blue diamonds, respectively. 
The figure shows that RCC relaxes ips to feed packets to WSAF table at the speed of 112 kips (thousand ips), which corresponds to 12\% regulation rate.
\ourFR{} effectively regulated flows to pass only 1.02\% with 128KB DRAM memory. As results, \ourFR{} has sufficient margin, while RCC does not have as can be seen in Fig.~\ref{fig-relax}.

\begin{figure}
	\centering
	\subfigure{\includegraphics[width=0.20\textwidth]{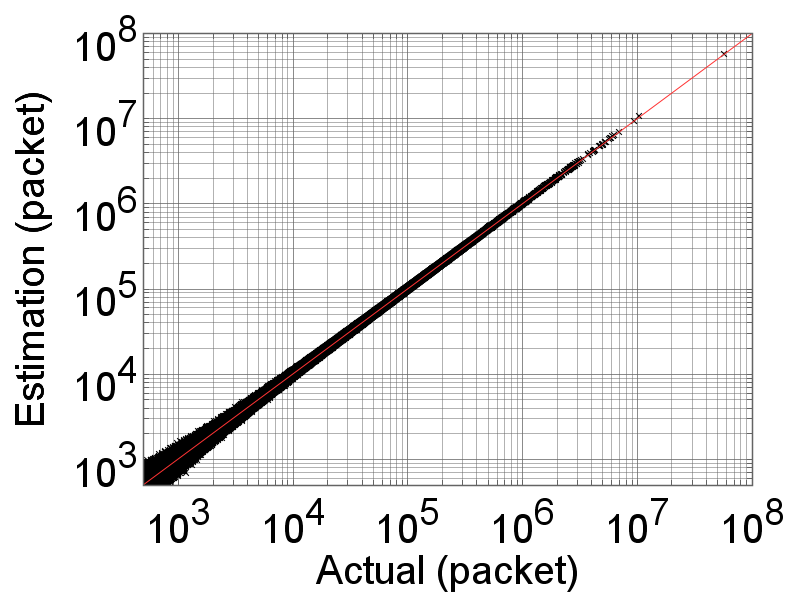}}
	\subfigure{\includegraphics[width=0.20\textwidth]{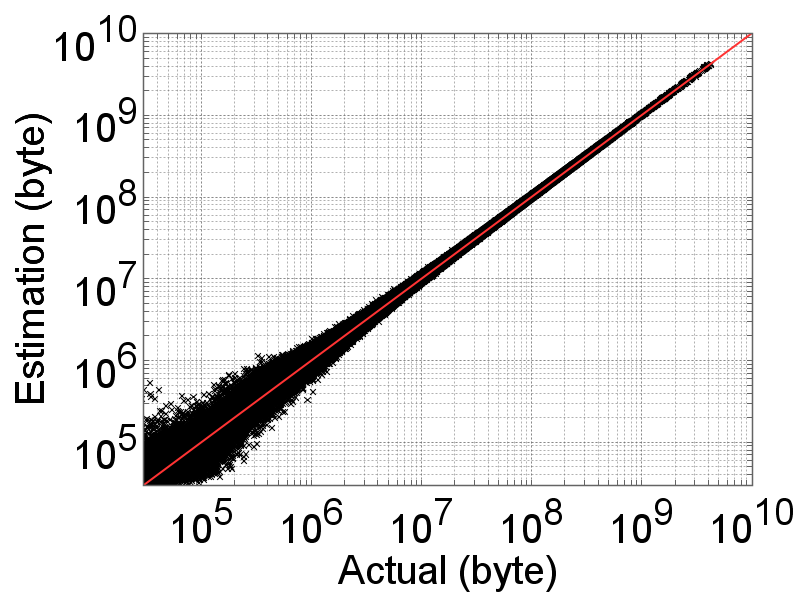}}\vspace{-2mm}
	\caption{Accuracy of packet counting (left) and byte counting (right).}\vspace{-3mm}
	\label{fig-accuracy}
\end{figure}
\begin{figure}
	\centering
	\subfigure{\includegraphics[width=0.20\textwidth]{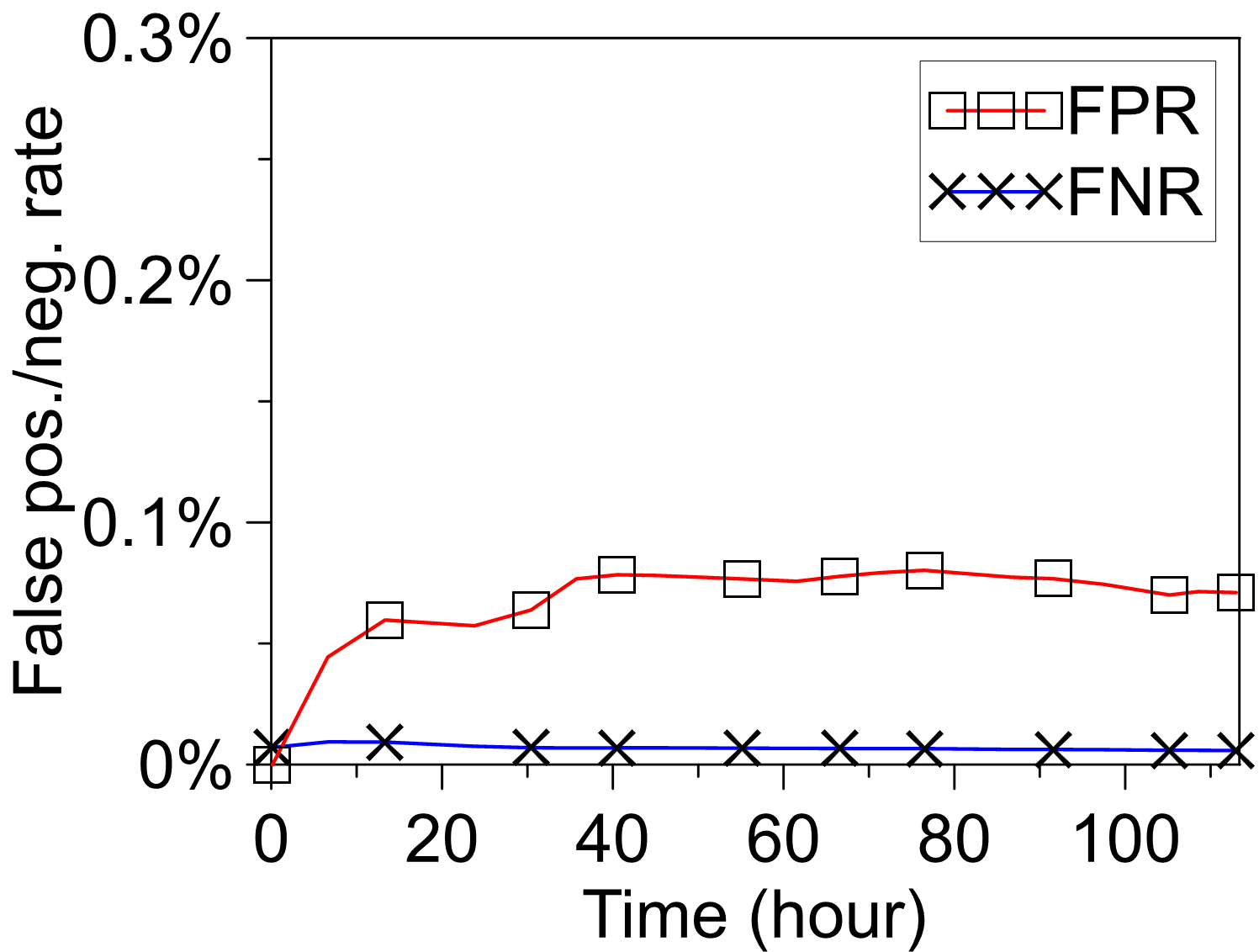}}
	\subfigure{\includegraphics[width=0.20\textwidth]{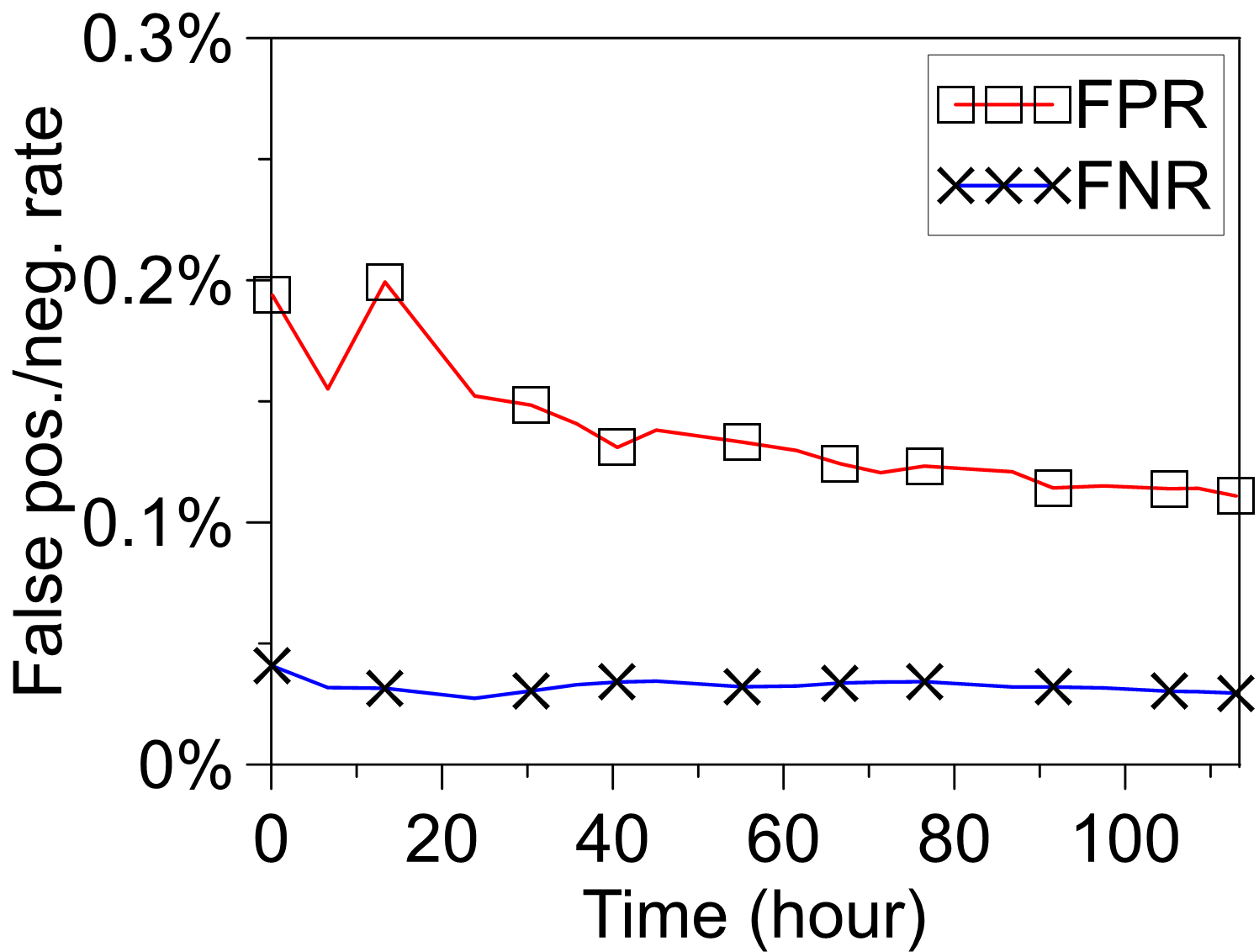}}\vspace{-3mm}
	\caption{False positive and false negative rates of packet heavy hitter detection (left) and byte volume heavy hitter detection (right).} \vspace{-3mm}
	\label{fig-hh}
\end{figure}

\BfPara{Monitoring in the wild}
We implemented our FlowRegulator in an off-the-shelf device and measured up-link traffics (1 Gbps bandwidth) at the backbone gateway (Juniper EX9208 switch) of our campus for 113 hours in total.
During 113 hours, 9.1 billion packets of 122.3 billion L4 flows were measured simultaneously both in packets and in bytes.

\BfPara{Accuracy} \ours{} used 128KB for sketch, and 33MB for the WSAF table. Sketches and WSAF table are all in DRAM.
Fig.~\ref{fig-accuracy} shows the estimation accuracy by standard error for the real-world experiment. 
For packet counting, we report 0.54\% standard error over 350 flows of which size is 1000K+, 
1.61\% over 11,047 flows for 100K packets, 3.46\% over 104292 flows for 10K+ packets.
For byte counting, we report 0.63\% over 414 flows of which byte size is 1G+,
1.74\% over 12,125 flows of 100MB+, 3.65\% over 107,726 flows of 10MB+.
This accuracy matches the accuracy observed in the lab experiment with the CAIDA dataset.

\BfPara{Heavy hitter detection}
Fig.~\ref{fig-hh} shows \ours{}'s heavy hitter detection accuracy in terms of false positive/negative rate.
Owing to \ours{} capability of counting both in packets and in  bytes, it can detect both packet heavy hitters
and byte heavy hitters. False negative rates in both cases are negligible, and the false positive rates of 
packet/byte heavy hitters are less than 0.1\% and 0.2\%, respectively.

\section{Conclusion} \label{sec:conclusion}
In this work, we have developed \ours{} for instant flow monitoring. Our approach is different from conventional measurement frameworks by introducing a new notion of very large In-DRAM working set of active flows. In the future work, we plan to demonstrate \ours{}'s performance and feasibility through an extensive analyses.

\BfPara{Acknowledgement} This work was supported by NRF grant number 2016K1A1A2912757 (Global Research Lab Initiative).

\vspace{-2mm}
\bibliographystyle{abbrv} 
\bibliography{ref}
\end{document}